\documentclass[twocolumn,prl,showpacs]{revtex4}

\usepackage{graphicx}
\usepackage{rotating}
\usepackage{amsmath}
\usepackage{amsfonts}
\usepackage{amssymb}
\usepackage{enumerate}
\usepackage{longtable}
\usepackage{dcolumn}
\usepackage{bm}

\begin{document}

\newcommand{\be}{\begin{equation}}
\newcommand{\ee}{\end{equation}}
\newcommand{\bn}{\begin{eqnarray}}
\newcommand{\en}{\end{eqnarray}}

\title{Orbital Order, Stripe Phases and Mott Transition in a
Planar Model for Manganites}

\author{M.S. Laad$^1$ and L. Craco$^2$}
\affiliation{$^1$Max-Planck-Institut f\"ur Physik komplexer Systeme, 
01187 Dresden, Germany \\ 
$^2$Max-Planck-Institut f\"ur Chemische Physik fester Stoffe, 
01187 Dresden, Germany} 
\date{\rm\today}

\begin{abstract}
Understanding orbital ordered (OO) Mott insulating states lies at the heart 
of a consistent resolution of the colossal magneto-resistance (CMR) observed 
in manganites, where its melting induces a low-$T$ insulator-metal transition 
for $0.25 \le x\le 0.45$. Motivated thereby, we study the OO states in a 
planar model for bilayer manganites using DMFT and finite-size 
diagonalisation methods. We derive the correct OO ground states observed 
in manganites for $x=0,\frac{1}{2},\frac{2}{3},\frac{3}{4}$ in exact 
agreement with observations, including the charge-orbital-magnetic ordered 
stripe phases for $x>\frac{1}{2}$. These OO states are {\it exactly} 
shown to be associated with an ``alloy'' ordering of the 
$d_{3x^{2}-r^{2}}/d_{3y^{2}-r^{2}}$ orbitals on each $Mn^{3+}$ site. 
\end{abstract}
     

\maketitle

Colossal magnetoresistance (CMR) materials have received much
attention~\cite{[1],[2]}, due to their extreme sensitivity to minute 
perturbations~\cite{[3]}.  The parent (cubic perovskite) materials are 
Mott-Hubbard insulators with $G$-type (AF) orbital order of 
$d_{3x^{2}-r^{2}}/d_{3y^{2}-r^{2}}$ orbitals and $A$-type AF spin 
order~\cite{[3]}.  Upon hole doping, $x$ (divalent ion substitution) in 
La$_{1-x}$Ca$_x$MnO$_3$, for example, they evolve through ferromagnetic, 
orbital ordered (OO) Mott insulators with unusual properties~\cite{[4]}, 
to a ferromagnetic metal (FM) at low-$T$.  A transition to a paramagnetic 
insulator (PI), dependent upon cation-dopant type, is seen for 
$T>T_{c}$~\cite{[3]}. A small magnetic field suppresses this I-M transition, 
leading to CMR. These phenomena are also seen in bilayer manganites. Further, 
more ``strange'' OO states are found in ``overdoped'' (with $Ca$) manganites.  
The half-doped manganites show a charge (C), orbital and AF order that is 
very sensitive to small perturbations~\cite{[5]} ($H_{ext}=5-7~T$ gives 
a ferromagnetic metal with no CO/OO).  The ``overdoped'' manganites with 
$x=\frac{1}{2},\frac{2}{3},\frac{3}{4},\frac{4}{5}$ show extremely stable 
pairs of $Mn^{3+}O_{6}$ Jahn-Teller distorted stripes having periods 
between $2-5 a$ ($a$=unit cell length); for other values of $x$, a mixture 
of the two adjacent commensurate configurations is found~\cite{[6]}. 
For $x=1$, ${\rm CaMnO_3}$ is again an AF ($S=\frac{3}{2},t_{2g}$) Mott 
insulator. Finally, the correlated nature of manganites is shown by dynamical 
spectral weight transfer (SWT) over large energy scales $O(4.0~eV)$ in 
various~\cite{[7],[8],[9]} studies as a function of $x,T,B_{ext}$-this can 
only result from strong electronic correlations. The importance of the 
Jahn-Teller (JT) coupling is evidenced by the large isotope 
effects~\cite{[10]} and by I-M transitions driven by 
$O^{18}\rightarrow O^{16}$ isotope substitution~\cite{[11]} (see, 
however, Ref.~\cite{[12]}, where the JT coupling is argued to be much 
weaker than in~\cite{[10],[11]}). Thus, understanding CMR is inextricably 
linked to understanding how these strongly coupled orbital-spin-charge 
correlations are modified by small perturbations as a function of $x$. 
A unified description of these unusual observations in one picture is a 
formidable challenge for theory.

The CMR problem has been extensively tackled in literature~\cite{[13],[14]} 
using a variety of numerical and analytic (QMC and $D=\infty$) methods, 
for double exchange (DE) models, with/without Jahn-Teller phonons, as 
well as with strong multi-orbital Coulomb interactions with static/dynamic 
JT phonons~\cite{[14]}.  For OO states, the full multi-orbital Hubbard 
model has been studied by mapping it to a Kugel-Khomskii (KK) 
model~\cite{[15]}. However, a controlled treatment (semiclassical 
analysis~\cite{[16]} indicates an order-by-disorder mechanism) is hard: 
even the type of order is unclear there, and the results sensitively 
depend on the approximations used~\cite{[17]}.  

Here, we take the first step to study the OO, Mott insulating phases 
observed in CMR manganites within a $2D$, multi-orbital Hubbard model 
incorporating the above-mentioned strongly coupled correlations.  Our
conclusions apply, with small additional modifications (to be treated 
separately) to bilayer manganites.  We show that a $2D$ model suffices 
to capture the correct OO states observed as a function of doping, $x$, 
and leave the full $3D$ problem for a separate work. Going beyond previous 
studies~\cite{[14],[17]}, we show how incorporation of the realistic 
structure of a single $MnO_{4}$ layer explicitly in the one-electron 
hopping integrals introduces new, unanticipated features, making a 
qualitative difference to the physical results for all $x$. Further, we 
show how the ``strange'' stripe-ordered phases in the global phase diagram 
are naturally rationalised from our effective model. 
  
We start with a model that explicitly includes orbital degeneracy of the 
$e_{g}$ orbitals in manganites~\cite{[2]},

\bn
\nonumber
H &=& - \sum_{<ij>a,b}t_{ij}^{ab}(a_{i\sigma}^{\dag}b_{j\sigma}+h.c) 
+ U\sum_{i,\beta =a,b}n_{i\beta\uparrow}n_{i\beta\downarrow} 
\\ \nonumber
&+& U'\sum_{i\sigma\sigma'}n_{ia\sigma}n_{ib\sigma'} 
- J_{H}\sum_{i\sigma\sigma'}{\bf S_{i}^{c} \cdot {\bf \sigma_{i}}}
(a_{i\sigma}^{\dag}a_{i\sigma'}+b_{i\sigma}^{\dag}b_{i\sigma'}) \\
&+& H_{JT} \;,
\label{HH}
\en
where the $a$ and $b$ are fermion annihilation operators in the doubly 
degenerate $e_{g}$ orbitals, $t_{ij}^{ab}$ 
($a,b=d_{3x^{2}-r^{2}},d_{3y^{2}-r^{2}}$) 
is a 2$\times $2 matrix in orbital space incorporating realistic features 
of the basic $Mn-O$ perovskite structure~\cite{[20]}. $U,U'$ are the on-site, 
intra- and inter-orbital Hubbard interactions, and $J_{H}$ is the (strong) 
Hund's rule coupling giving rise to the FM state as in the usual DE model. 
Polaronic effects are described by $H_{JT}$ (see below).  

At strong coupling, setting $U,J_{H}>>t$ gives the following effective 
Hamiltonian, $H_{0}=-\sum_{ij,a,b,\mu} t_{\mu}^{ab}\gamma_{ij}({\bf S})
(a_{i}^{\dag}b_{j}+h.c)$ with $\mu=x,y$.
Here, $t_{x}^{ab}=\frac{t}{4}[3,\sqrt{3},\sqrt{3},1]$ and 
$t_{y}^{ab}=\frac{t}{4}[3,-\sqrt{3},-\sqrt{3},1]$ define the one-electron 
hopping matrix for a single manganite layer. We now turn on $U'$. One is 
effectively dealing with spinless fermions, but now with an orbital index. 
Clearly, this model ($U'=0$) cannot access the interplay between magnetism 
and OO in manganites. With $U'$ and the JT coupling terms, $H$ becomes 

\be
H_{eff} = H_{0} 
+ U'\sum_{i,a\ne b}n_{ia}n_{ib} + H_{JT} \;,
\ee
where $\gamma_{ij}({\bf S})$ is the usual DE projection factor~\cite{[3]}.  

Transform to new variables, 
$c_{{\bf \alpha}\uparrow}=(a+(-1)^{\bf \alpha}\sqrt{3}b)/\sqrt{2}$, 
$c_{{\bf \alpha}\downarrow}=((-1)^{\bf \alpha}\sqrt{3}a-b)/\sqrt{2}$  with
$(-1)^{\bf \alpha} \equiv +1~(\alpha || x)$ and $\equiv -1~(\alpha || y)$.
The $c_{\alpha\sigma}$ transform exactly like 
$d_{3x^{2}-r^{2}}(\uparrow),d_{3y^{2}-r^{2}}(\downarrow)$.
This {\it exactly} yields a Falicov-Kimball model (FKM) where only the 
$c_{{\bf \alpha}\uparrow}$ hop; the $c_{{\bf \alpha}\downarrow}$ are 
strictly immobile as long as no JT distortions are included. Thus,

\bn
\nonumber
H_{eff} &=& -\sum_{<ij>,\alpha} t\gamma_{ij}({\bf S})
(c_{i{\bf \alpha}\uparrow}^{\dag}c_{j{\bf \alpha}\uparrow} + h.c)
+ U'\sum_{i,\alpha}n_{i\alpha\uparrow}n_{i\alpha\downarrow} \\ 
&+& H_{JT} \equiv H_{FKM} + H_{JT} \;,
\en
reflecting the correlation between the magnetic and orbital degrees of 
freedom described above.  

In orbital space, the JT coupling corresponds to addition of external 
fields~\cite{[20]}, 
$H_{JT}=Q_{2}\sum_{i}(n_{ia}-n_{ib})+ Q_{3}\sum_{i}(a_{i}^{\dag}b_{i}+h.c)$.  
In the rotated basis, this is, 

\be
H_{JT}=Q_{++}\sum_{i,\alpha}(n_{i{\bf \alpha}\uparrow}
-n_{i{\bf \alpha}\downarrow})
+ Q_{+-}\sum_{i,\alpha}(c_{i{\bf \alpha}\uparrow}^{\dag}
c_{i{\bf \alpha}\downarrow}+h.c) \;,
\ee
where $Q_{++}=((-1)^{\alpha}\sqrt{3}Q_{2}-Q_{3})/2$ and 
$Q_{+-}=(Q_{2}+(-1)^{\alpha}\sqrt{3}Q_{3})/2$ are staggered JT distortions 
which follow the orbital (electronic) variables. So $H_{eff}=H_{FKM}+H_{JT}$ 
is a FKM with a local, staggered hybridisation between the 
$c_{{\bf \alpha}\uparrow},c_{{\bf \alpha}\downarrow}$ at each site.
Inclusion of finite phonon frequency ($M\Omega^{2}(Q_{2}^{2}+Q_{3}^{2})/2$) and
intersite phonon coupling terms is required in a full analysis: we have not 
done this here.

\begin{figure}[t]
\includegraphics[width=\columnwidth]{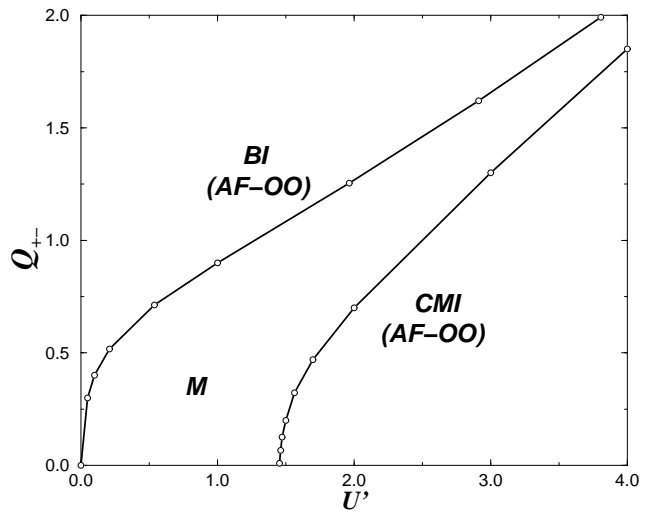}
\caption{Phase diagram for the transformed multi-orbital model with $Q_{++}=0$
at half-filling.  The charge Mott (CMI) and the band (BI) insulators, both 
with anti-ferro-orbital order (AF-OO) are separated by an incoherent, 
pseudogapped metal (M) phase. }
\label{fig2}
\end{figure}

For a 
half-filled band of spinless fermions, the exact solution of $H$ in $D=2$ 
implies an anti-ferro orbital order of $d_{3x^{2}-r^{2}},d_{3y^{2}-r^{2}}$, 
exactly as required~\cite{[21]}. Such a FKM has been employed 
earlier~\cite{[14],[17]} for mangnanites, but  
$c_{\alpha\uparrow}=d_{x^{2}-y^{2}}, 
c_{\alpha\downarrow}=d_{3z^{2}-r^{2}}$ there.
This would lead to an AFOO of $d_{x^{2}-y^{2}}/d_{3z^{2}-r^{2}}$, at variance 
with observations. Here, such a FKM follows {\it exactly} from the realistic 
hopping structure.  Moreover, the AF-OO (Mott insulating, see below) state
is driven by large $U'$, in contrast with band-based scenarios.
We note that Yamasaki {\it et al.}~\cite{ya} have derived an AF-OO Mott 
insulator for cubic ${\rm LaMnO_3}$ (with $x=0$) using LDA+DMFT. Our work is 
thus complementary to theirs for $x=0$, but goes much further, permitting 
us to study the ``exotic'' OO states for $x\ge \frac{1}{2}$ as well (see 
below).  Moreover, given our effective FK mapping~\cite{[23]}, the OO 
state(s) are readily understood in terms of an ``alloy'' ordering of 
$d_{3x^{2}-r^{2}},d_{3y^{2}-r^{2}}$ orbitals at {\it each} $Mn$ site. 

We now solve $H_{eff}=H_{FKM}+H_{JT}$ in $d=\infty$. As shown 
earlier~\cite{[21]}, DMFT works surprisingly well for the $2D$ FKM. 
The FKM with/without $Q_{+-}$ has an almost exact solution in 
$D=\infty$~\cite{[22]}. The formalism is essentially the same as that 
used previously, and gives very good agreement with QMC results for the 
same model~\cite{qi}. Keeping $U'/t$ fixed and large, phase transitions 
from the Mott insulator with AF-OO to correlated (incoherent) metal with 
no OO, to a correlation-asssisted band insulator, again with AF-OO, occur: 
this is indeed borne out in the $D=\infty$ solution, as shown in 
Fig.~\ref{fig2}. Given that $U'$ is much larger than $Q_{++,+-}$ in $H$ 
above, we conclude that manganites fall into the CMI class with AF-OO, 
and that the JT terms lead to additional stabilization of both.  Finally, 
DMFT gives the full, correlated spectral functions of the model for 
arbitrary parameter values and band-fillings, at a very modest numerical 
cost.  This allows us to study the filling driven Mott transition from an 
AF-OO Mott insulator to an incoherent metal (see below).  

\begin{figure}[t]
\includegraphics[width=\columnwidth]{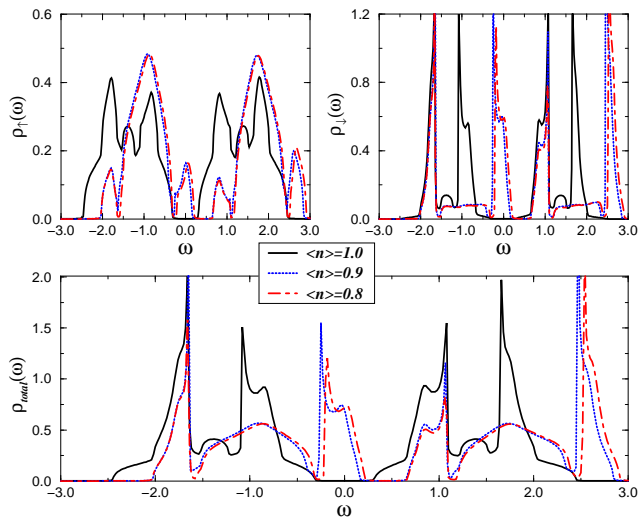}
\caption{(Color online) Partial orbital-resolved (top panels) and the total 
many-body DOS for the AF-OO phase of $H_{eff}$ (see text) 
within DMFT for $U'=2.6$~eV, and {\it staggered} $Q_{++}=0.3$~eV, 
$Q_{+-}=0.4$~eV, for various band-fillings. Off-diagonal components of 
the spectral fuction are not shown. For $\langle n\rangle=0.9,0.8$, the 
low-energy pseudogap at $\omega=0$ is clearly seen in the DOS.}
\label{fig3}
\end{figure}

The relevant DMFT equations were derived earlier~\cite{[22]}, so we do not 
repeat them here. Since the JT terms are staggered, but bilinear in the 
$e_{g}$ basis, they are easily incorporated into the earlier DMFT structure.
The Green function is now a $(2 \times 2)$ matrix in orbital space. The 
staggered, JT ``external field'' terms imply an averaging over their
 orientations, which is carried out within the DMFT equations to yield 
the DOS. We choose $U'=2.6$ eV, $Q_{++}=0.3$ eV, $Q_{+-}=0.4$ eV as model 
parameters along with a non-interacting DOS for the 2D square lattice with 
bandwidth, $W=2.0$ eV and variable band-filling, $n=(1-x)$, in the DMFT 
solution. For $n=1$, (see Fig.~\ref{fig3}) we obtain an AFOO Mott insulator. 
This is obtained from the computed value of 
$D_{1\alpha}=(-1)^{\alpha}\langle(c_{i\alpha\uparrow}^{\dag}c_{i\alpha\downarrow}+h.c)\rangle=C(\frac{U'}{W},Q_{++,+-})=0.07$ and $D_{2\alpha}=(-1)^{\alpha}\langle(n_{i\alpha\uparrow}-n_{i\alpha\downarrow})\rangle=C'(\frac{U'}{W},Q_{++,+-})=0.05$ (not shown), obtained directly from $D_{2\alpha}=-\frac{1}{\pi}\int \sigma Im G_{\alpha\sigma}(\omega)d\omega$ and $D_{1\alpha}=-\frac{1}{\pi}\int Im G_{\alpha\uparrow\downarrow}(\omega) d\omega$ from the DMFT equations.  Away from 
$n=1$, the DMFT equations have to be supplemented with the Friedel-Luttinger 
sum rule, $\langle n \rangle=-\frac{1}{\pi}\int_{-\infty}^{E_{F}}\sum_{\alpha,\sigma} Im G_{\alpha\sigma}(\omega)d\omega$. 
This is computed self-consistently within the DMFT.

For $\langle n\rangle=0.9,~0.8$, we obtain an {\it incoherent}, pseudogapped, 
metallic state (see Fig.~\ref{fig3}) with a sharp reduction of local 
anti-ferro orbital (AFO) correlations ($D_{1\alpha}=0.009$). Thus, appearance 
of the doping-driven (FM) metallic state is intimately linked to the 
``melting'' of local anti-ferro orbital correlations of the Mott insulator 
with $x$. The non-FL character of the FM contrasts with what is expected in 
the FKM with uniform hybridisation ($V=Q_{+-}$ in the usual FKM with 
hybridization), where a correlated FL metal is obtained whenever $V$ is 
relevant~\cite{[MH]}. In our model, the staggered ``fields'' $Q_{++,+-}$ 
produce a low-energy pseudogap, suppressing FL coherence. Chemical disorder 
will further reinforce incoherence~\cite{[24]}.  Given the $d$-wave character 
of the staggered JT terms (note that both $Q_{++,+-}$ have components that 
change sign under a $\pi/2$ rotation in $xy$ plane), as well as the (more
important) fact that $d$-wave ground states are obtained near half-filling 
in a Hubbard-like (FKM) model~\cite{[GK]}, we predict that this incoherent 
FM-metal phase will exhibit a $d$-wave pseudogap. 

\begin{figure}[t]
\includegraphics[width=\columnwidth]{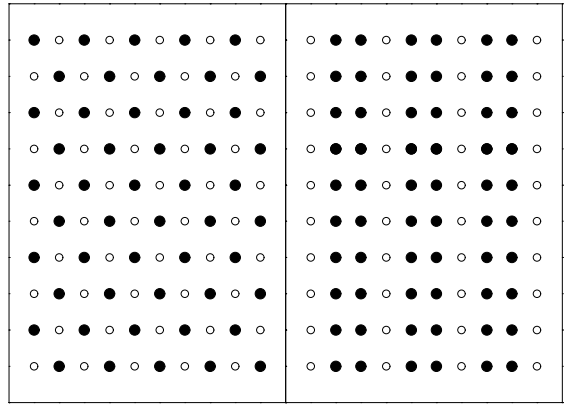}
\caption{Two different charge-orbital ordered (COO) ground states of the 
effective Falicov-Kimball model for $x=\frac{1}{2}$ (left) and 
$x=\frac{2}{3}$ (right).  These exactly correspond to the COO 
states observed in manganites for these $x$ values~\cite{[2],[6]}.}
\label{fig4}
\end{figure}

In contrast to earlier FKM work~\cite{[14],[17]}, however, the ordered, 
insulating phases in un(doped) manganites arise naturally from our model. 
The checkerboard order of $d_{3x^{2}-r^{2}},d_{3y^{2}-r^{2}}$ corresponds 
to an AF-OO insulator. The exotic bi-stripe states too are 
naturally predicted from the analysis of our FKM. In the insulating phases, 
the ``hybridisation''($Q_{+-}$) is irrelevant, and the resulting FKM 
rigorously undergoes phase separation into hole-rich (orbital disordered) 
and hole-poor (orbitally ordered)  phases, as shown by Freericks 
{\it et al.}~\cite{[23]} by minimizing the total energy for various $x$. 
We have repeated their analysis for various $x\ge 0.5$. For 
$x=\frac{1}{2},\frac{2}{3},\frac{3}{4},\frac{4}{5}$, we obtain stripe 
phases with periods $2,3,4,5$, as observed by Mori {\it et al.}~\cite{[6]} 
using electron diffraction.  In Fig.~\ref{fig4}, we show only the OO 
ground states for $x=\frac{1}{2},\frac{2}{3}$; these exactly correspond 
to those observed in manganites for these hole dopings.

Given that $Mn^{3+,4+}$ correspond to one/zero $e_{g}$ electron on each 
$Mn$ site, the $2D$ model automatically has charge order (CO) of the 
correct types for these values of $x$.  Also, the stripe OO of pairs of 
$Mn^{3+}O_{6}$ (distorted) octahedra automatically corresponds to a 
bi-stripe charge-order (CO) of $e_{g}$ electrons with the periodicity 
determined by $x$~\cite{[2]}.  Given the bi-stripe OO states, 
Goodenough-Kanamori-Anderson rules directly imply that intersite 
interactions between the ``core'' $t_{2g}$ spins ($S=3/2$) will lead to
AF-coupled ladders ($Mn^{3+}$) separated by strips of JT-undistorted 
($Mn^{4+}$) regions.  Given suppression of $e_{g}$ hopping in an AF 
``background'', these stripe states will be insulators,  
as observed~\cite{[1],[2]}. These states will be further stabilised upon 
inclusion of JT terms and longer range elastic interactions. 
                                           
This fully corresponds to observations in bilayer manganites for 
$x>0.5$~\cite{[2]}. Thus, stripe states in ``overdoped'' CMR {\it exactly} 
result from an ``alloy'' ordering of a binary alloy of $Mn^{3+}(S=2,d^{4})$ 
and $Mn^{4+} (S=\frac{3}{2},d^{3})$ orbitals with 
$d_{3x^{2}-r^{2}},d_{3y^{2}-r^{2}}$ 
symmetry. Phase separation/stripe phases have long been studied using the 
FKM (binary alloy disorder model) in alloy physics~\cite{[25]}. Here, we 
show how these phenomena in manganites arise from strong, multi-orbital, 
{\it electronic} correlations, which are now exactly representable as a 
binary alloy model. Since OO states spontaneously break {\it discrete}, 
Ising symmetries of $H$ (Eq.~(\ref{HH})), the link to alloy ordering 
(described within an Ising model framework~\cite{[25]}) is readily 
apparent. Recently~\cite{[AT]}, OO phases in a $3D$ model were derived 
within a static Hartree-Fock approximation. In future, we will make 
contact with these results.   

To conclude, we have shown how consideration of the actual multi-orbital 
structure of the hopping matrix in the $e_{g}$ sector within a multi-orbital 
correlated model results in an understanding of the various OO insulating 
phases observed in CMR manganites, especially in bilayer cases, as a function 
of $x$. These are now understood simply as an ``alloy'' ordering of 
$d_{3x^{2}-r^{2}},d_{3y^{2}-r^{2}}$ orbitals, driven predominantly by the 
inter-orbital correlations ($U'$). Our study shows that OO in overdoped
($x>0.5$) manganites 
need not imply very strong JT coupling, in agreement with~\cite{[12]}: 
by itself, $U'$ leads directly to such phases as a function of $x$. A 
moderate JT distortion will further stabilise these ordered phases. Within 
multi-orbital DMFT, we have shown how an AFOO/F Mott insulator turns into 
a correlated, incoherent, ferromagnetic ``bad metal'' upon hole doping. 
This goes hand-in-hand with a drop in local AFO correlations. These results 
are fully consistent with indications from a host of experiments probing 
various phases of doped bilayer manganites. Interestingly, planar nickelates 
are also modelled by a similar Hamiltonian, and our work also naturally 
explains the OO/stripe phases observed there~\cite{[26]}. We expect our 
analysis to be broadly applicable to a variety of TMO systems showing a 
variety of OO/magnetic ground states as a function of a suitable ``tuning 
parameter''.

M.S.L thanks Prof. P. Fulde for advice and support at the MPIPKS, Dresden.
L.C. thanks the Emmy Noether-Programm of the DFG for financial support.


\begin{thebibliography}{28}  
   
\bibitem{[1]} see ``Colossal Magnetoresistance Manganites'', ed. Y. Tokura
(Gordon and Breach, New York, 2000).

\bibitem{[2]} see ``Nanoscale Phase Separation in Manganites'', by E. Dagotto
(Springer Verlag, NY and Heidelberg, 2002), and references therein.

\bibitem{[3]} S. Ishihara, J. Inoue, and S. Maekawa,
Phys. Rev. B {\bf 55}, 8280 (1997).
 
\bibitem{[4]} S. Uhlenbruck {\it et al.}, 
Phys. Rev. Lett. {\bf 82}, 185 (1999).

\bibitem{[5]} Y. Tomioka {\it et al.}, 
Phys. Rev. Lett. {\bf 74}, 5108 (1995).

\bibitem{[6]} S. Mori, C. H. Chen and S.-W. Cheong, Nature {\bf 392}, 
473 (1998).

\bibitem{[7]} Y. Okimoto {\it et al.}, 
Phys. Rev. Lett. {\bf 75}, 109 (1995).

\bibitem{[8]} D. Dessau {\it et al.}, 
Phys. Rev. Lett. {\bf 81}, 192 (1998).

\bibitem{[9]} J. Simpson {\it et al.}, 
Phys. Rev. B {\bf 60}, R16263 (1999).

\bibitem{[10]} G.-M. Zhao {\it et al.}, 
Nature, (London) {\bf 381}, 676 (1996).

\bibitem{[11]} N. A. Babushkina {\it et al.}, 
Nature {\bf 391}, 159 (1998).

\bibitem{[12]} J. C. Loudon {\it et al.}, 
Phys. Rev. Lett. {\bf 94}, 097202 (2005).

\bibitem{[13]} N. Furukawa, J. Phy. Soc. Jpn. {\bf 64}, 2734 (1995); see 
also Ref.~\cite{[2]}.

\bibitem{[14]} M. S. Laad, L. Craco, and E. M\"uller-Hartmann, 
Phys. Rev. B {\bf 63}, 214419 (2001);
T. V. Ramakrishnan {\it et al.}, in ``Colossal Magnetoresistance Manganites'',
ed T. Chatterji, Kluwer Acad. Publ, Netherlands (2003).

\bibitem{[15]} K. Kugel and D. I. Khomskii, Sov. Phys.-JETP {\bf 37}, 
725 (1973).

\bibitem{[16]} G. Khaliullin and V. Oudovenko, Phys. Rev. B {\bf 56}, 
R14243 (1997); L. F. Feiner, A. M. Oles, and J. Zaanen, J. Phys. Cond. 
Matter. {\bf 10}, L555 (1998). These reach opposite conclusions, which 
sensitively depend on the decouplings used.

\bibitem{[17]} V. Ferrari, M.J. Rozenberg, and R. Weht, 
Mod. Phys. Lett. B {\bf 15}, 1031 (2001); also see Ref.~\cite{[14]}.

\bibitem{[20]} T. Kennedy and E. H. Lieb, Physica {\bf 138A}, 320 (1986); 
ibid E. H. Lieb, Physica {\bf 140A}, 240 (1986).

\bibitem{[21]} J. Freericks and V. Zlati\'c, Rev. Mod. Phys. {\bf 75}, 
1333 (2003).

\bibitem{ya} A. Yamasaki {\it et al.}, 
Phys. Rev. Lett. {\bf 96}, 166401 (2006).

\bibitem{[23]} R. Lemanski, J. K. Freericks, and G. Banach, 
J. Stat. Phys. {\bf 116}, 699 (2004).

\bibitem{[22]} L. Craco, Phys. Rev. B {\bf 59}, 14837 (1999).

\bibitem{qi} Q. Si {\it et al.}, 
Phys. Rev. Lett. {\bf 72}, 2761 (1994).

\bibitem{[MH]} E. M\"uller-Hartmann, T. V. Ramakrishnan, and G. Toulouse,
Phys. Rev. B {\bf 3}, 1102 (1971).

\bibitem{[GK]} T. Stanescu and G. Kotliar, Phys. Rev. B {\bf 74}, 125110 
(2006).

\bibitem{[24]} M. S. Laad, L. Craco, and E. M\"uller-Hartmann, 
Phys. Rev. B {\bf 64}, 195114 (2001).

\bibitem{[25]} D. de Fontaine, in ``Solid State Physics'', eds. 
H. Ehrenreich {\it et al.}, Vol 34, pg. 73 (Academic Press, NY, 1979).

\bibitem{[AT]} A. Taraphder, J. Phys. Condens. Matt. {\bf 19}, 125218 (2007). 

\bibitem{[26]} T. Hotta and E. Dagotto, Phys. Rev. Lett. {\bf 92}, 227201 
(2004).

\end{thebibliography}
\end{document}